\documentclass[12pt,preprint]{aastex}

\usepackage{color}                    

\def \apj {ApJ}

\def \apjl {ApJ}

\def \aap {A\&A}

\begin{document}

\title{Spectropolarimetric inversions of the \ion{Ca}{2} 8498 \AA\ and 8542 \AA\ lines in the quiet Sun}

\author{A. Pietarila\altaffilmark{1}, H. Socas-Navarro}
\affil{High Altitude Observatory, National Center for Atmospheric Research\altaffilmark{2}, 3080 Center Green, Boulder, CO 80301, USA\vbox{}}
\author{T. Bogdan}
\affil{Space Environment Center, National Oceanic and Atmospheric Administration, 325 Broadway, Boulder, CO 80305, USA\vbox{}}

\altaffiltext{1}{Institute of Theoretical Astrophysics, University of Oslo, P.O.Box 1029 Blindern, N-0315 Oslo, Norway}
\altaffiltext{2}{The National Center for Atmospheric Research (NCAR) is sponsored by the National Science Foundation.}

\begin{abstract}
We study non-LTE inversions of the Ca II infrared triplet lines as a tool for inferring physical properties of
the quiet Sun. The inversion code is successful in recovering the
temperature, velocity and longitudinal magnetic flux density in the
photosphere and chromosphere, but the height range where the
inversions are sensitive is limited, especially in the
chromosphere. We present results of inverting spectropolarimetric observations of the lines in a quiet Sun region. We find
three distinct ranges in chromospheric temperature: low
temperatures in the internetwork, high temperatures in the enhanced magnetic
network and intermediate temperatures associated with low magnetic
flux regions in the network. The differences between these regions become more pronounced with height as the plasma-$\beta$ decreases. These inversions support the picture of the chromosphere, especially close to the magnetic network, being highly inhomogeneous both in the vertical and horizontal directions.

\end{abstract}

\section{Introduction}

A transition from a high plasma-$\beta$ ($\beta=p_{gas}/p_{mag}$) to a
low-$\beta$ regime takes place in the chromosphere. With this
transition the nature of the dynamics
changes: in the high-$\beta$ regime the plasma dominates the magnetic fields whereas in the low-$\beta$ regime the opposite is true. Waves undergo mode mixing, reflection and refraction in the layer where plasma-$\beta \approx 1$. Because of the tight connection between the magnetic fields and dynamics it is crucial
to understand how the two are coupled in the quiet Sun chromosphere. Several
recent observational works have focused on this
(e.g., \citealt{Judge+others2001, McIntosh+Judge2001, McIntosh+Fleck+Judge2003, Jefferies+others2006} and \citealt{Vecchio+others2007}). All find a clear correlation between the
magnetic field topology and waves in the chromosphere. None of the
above mentioned works, however, directly addresses the magnetic field
in the chromosphere. Instead, proxies such as potential field
extrapolation or the existence of fibril-like structures in the
emission of chromospheric lines, are used.

In order to obtain physical parameters (instead of proxies) from spectropolarimetric
observations an inversion problem is
unavoidable. Since the radiative transfer equation for polarized light
is a set of coupled differential equations, and in the chromosphere
local thermodynamical equilibrium (LTE) is no longer valid, solving the
inverse problem is challenging. Furthermore, one is never certain that any
solution obtained is in fact the correct solution, instead we have
to rely on some {\it a priori} knowledge of the solar
atmosphere. Inversions are commonly used to to produce quantitative information of the photosphere. They are not yet as widely used for chromospheric observations. Significant progress in chromospheric inversions has been made for the
\ion{He}{1} 10830 \AA\ multiplet (e.g., \citealt{Lagg+others2004,
Centeno+others2005, TrujilloBueno+others2005, Sasso+Lagg+Solanki2006,
Merenda+others2006}) and \ion{Ca}{2} infrared (IR) triplet lines
(e.g., \citealt{Socas-Navarro+others2000b, Socas-Navarro2005a,
Socas-Navarro+others2006b}).

The magnetic structure of the quiet Sun chromosphere is still a matter
of debate, but there is no doubt that it plays an important role. The leakage of
low-frequency (i.e., below acoustic cut-off frequency) waves into the
chromosphere and the formation of spicules have been attributed
to the existence of inclined magnetic fields in the network
(e.g., \citealt{DePontieu+others2004, Jefferies+others2006,
Hansteen+others2006, DePontieu+others2007}). It is also suggested that
these waves may be related to the heating of the chromosphere
(\citealt{Jefferies+others2006}). Existence of canopy-like structures such as proposed by
\cite{Gabriel1976} is still an open debate. For example \cite{Vecchio+others2007} observe oscillatory patterns in the
chromosphere that are highly consistent with the canopy-scenario, while
models by e.g., \cite{Schrivjer+Title2003} and
\cite{Jendersie+Peter2006} show that the existence of a canopy-like
structure is disrupted by internetwork magnetic fields. The connection between magnetic fields, dynamics and thermal properties of the quiet Sun chromosphere appears complex. In this paper we extend the use of non-LTE inversions to the quiet Sun chromosphere. With non-LTE inversions we do not have to rely on proxies, such as line intensity, to infer physical quantities. 

\nocite{Pietarila+others2006}
\nocite{Pietarila+others2007a}

This paper is a continuation to Pietarila et al.\ (2006, hereafter P06) and Pietarila et al.\ (2007, hereafter P07) in which the use of \ion{Ca}{2} IR triplet lines as a diagnostic for chromospheric magnetism are studied. Here we focus on non-LTE inversions of the \ion{Ca}{2} IR
triplet lines, and what information can be extracted from them. In
section \ref{sec:code} the non-LTE inversion code used in this work is
described in some detail. Also the
sensitivity of the inversions is discussed. We demonstrate that the Ca lines
are sensitive only in a narrow height range in the chromosphere, and
that our inversion code cannot recover discontinuities in magnetic field inclination
or velocity (at least with the observables used). We also show that discontinuities can cause
complex patterns in Ca line Stokes $V$ asymmetries. Inversions of
quiet Sun observations are discussed in section \ref{sec:obs}. The
network is found to be more dynamic than the internetwork. Three different
regimes are found in chromospheric temperature: low temperatures
corresponding to the ``magnetic field free '' internetwork, high
temperatures in the enhanced magnetic network and an intermediate temperature
regime associated with low magnetic flux regions. The main
results are discussed (in the context of previous work) in section \ref{sec:disc}, and final remarks are made in section \ref{sec:conc}.

\section{Inversion code}

The non-LTE inversion code employed here has been described in detail in \cite{Socas-Navarro+RuizCobo+TrujilloBueno1998} and \cite{Socas-Navarro+TrujilloBueno+RuizCobo2000}. For completeness, we provide here a brief summary of its main features. The user needs to supply an initial model atmosphere, which consists of the run with standard optical depth ($\tau_{500}$) of temperature, microturbulence, line-of-sight velocity and magnetic field vector (strength, inclination and azimuth). In addition to these vector quantities, a model atmosphere includes the following single-valued parameters: electron pressure at the top of the atmosphere, macroturbulence and filling factor. The macroturbulence is used also to describe the spectral instrument profile. The filling factor is used to consider that the magnetic structure is smaller than the resolution element (or that there is some stray light in the spectrograph).

With the top boundary condition for the electron pressure and the run of temperature, the hydrostatic equilibrium equation is solved to obtain the run of electron pressure, gas pressure and density and the geometrical scale height. The code then uses a preconditioning strategy (\citealt{Rybicki+Hummer1991, Socas-Navarro+TrujilloBueno1997}) to solve the statistical equilibrium equations in that atmosphere and synthesizes the Stokes profiles emerging from the initial model as well as the response functions, i.e. the derivatives of the Stokes vector at a given wavelength with respect to the model parameters.

The derivatives enter a Levenberg-Marquardt scheme that determines the first-order correction needed to compensate for the misfit between the observed and the synthetic profiles. The correction for a given depth-dependent parameter (e.g., temperature) is obtained as a smooth function of $\tau_{500}$, depending on the number of {\it nodes} selected by the user for that parameter (for two nodes the code retrieves a correction that is linear with $\tau_{500}$, for three nodes a parabola, and for four or more nodes the correction is a cubic spline interpolation). The entire algorithm is iterated until it converges to a solution that minimizes the difference between the observed and synthetic profiles.

\label{sec:code}
\subsection{Sensitivity of the inversion code}
\label{sec:tests}

\label{sec:beta}

\begin{figure*}
\epsscale{.8}
\plotone{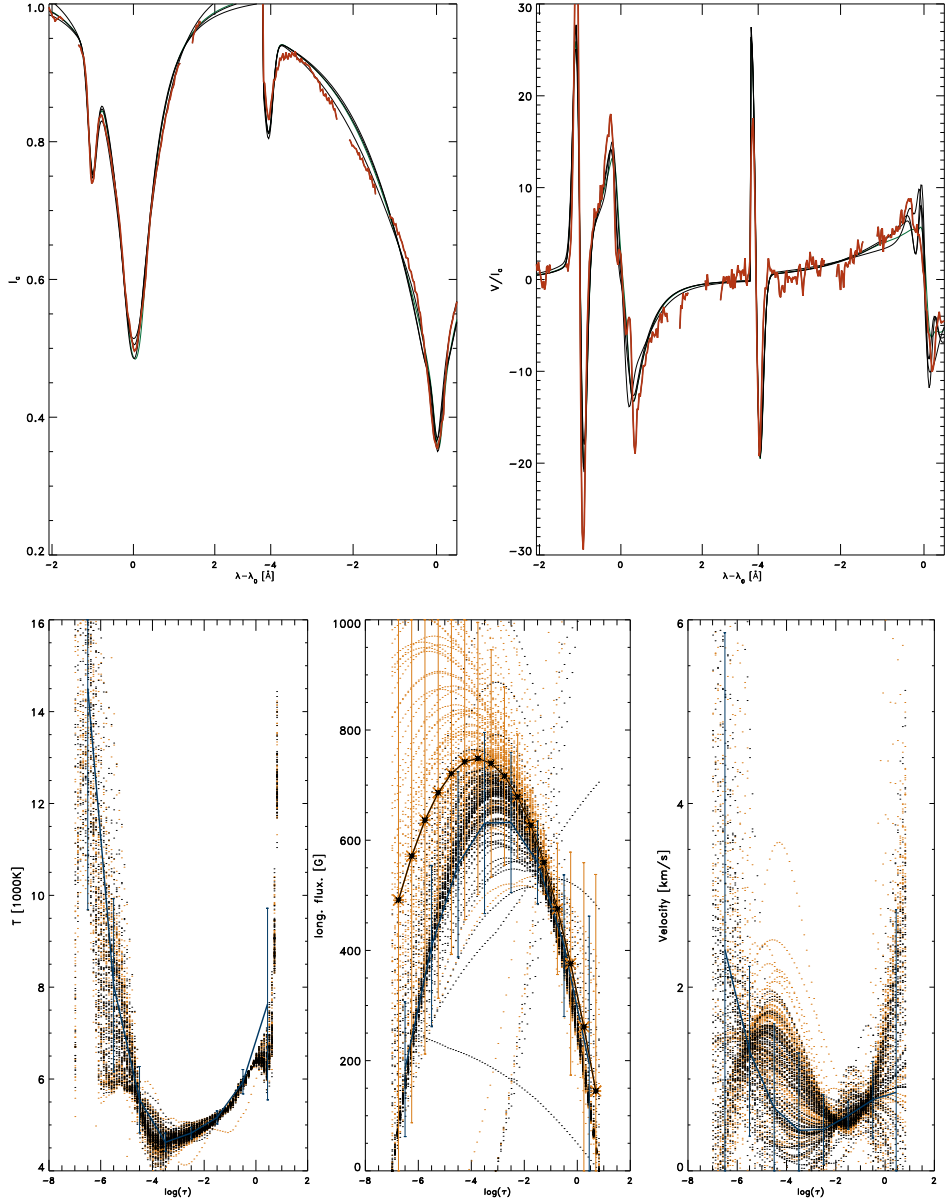}
\caption{ 
\label{fig:sensitivity}
Sensitivity of the lines in the inversions. Upper row: input Stokes $I$ and $V$ profiles are in red and fits to the profiles in black lines. Lower row: Temperature (left), longitudinal magnetic flux (middle) and velocity (right) as a function of optical depth. Black points show results for models with variation in magnetic field strength only, orange points for models with variation in both inclination and magnetic field strength. Blue and orange curves show the mean of the black and orange points, respectively, and the error bars show the standard deviation.}
\end{figure*}

The inversion code retrieves model atmospheres with the atmospheric
parameters as a function of optical depth, ranging from
$log(\tau)=[-6$, $0.8]$. The sensitivity of the spectral lines,
however, is limited to only a range of atmospheric depths. To identify
what the depths are we made the following numerical experiment: a set of
simultaneous Stokes $I$ and $V$ profiles of the 8498 \AA\ and 8542
\AA\ \ion{Ca}{2} lines and the two photospheric \ion{Fe}{1} lines,
8497 \AA\ and 8538 \AA\ (located in the wings of the Ca lines), were
inverted using 100 different initializations. The initial models were
chosen randomly from existing inversions of observations. To make
the test comparable to the observations discussed later, we do not
include Stokes $Q$ or $U$ profiles. Additionally, we force the magnetic field to be vertical, i.e., only the longitudinal magnetic field is considered. The lines are, as expected, sensitive to
only part of the atmosphere (fig. \ref{fig:sensitivity}). The spread in the output models is smaller in two regions: in the photosphere, where the Fe lines and the Ca line wings are
formed, and in the chromosphere, where the Ca line cores are formed. For
temperature the heights are $log(\tau) \approx 0$ and $log(\tau) \approx
-5$ , for magnetic flux $log(\tau) \approx -1$ and $log(\tau) \approx -5$
, and for velocity $log(\tau) \approx -1$ and $log(\tau) \approx -5.5$. Since
neither the Ca nor the Fe lines are formed in the region between
$log(\tau) \approx [-4, -2]$ the output models have a large spread
at these heights. We conclude that the inversion results at this
height range are not reliable, and that the retrieved atmospheric
values are reliable only in the fairly narrow height ranges.

In figure \ref{fig:sensitivity} are also shown the retrieved magnetic
field strengths with nodes in both the field strength (3 nodes) and
inclination (4 nodes). The spread of field strengths obtained is unreasonably
large, especially in the chromosphere. Since the inversion code has no
information of Stokes $Q$ and $U$, some ambiguity between the magnetic
field strength and inclination cannot be removed. Due to this, for
the observations we choose to force the inclination to zero and
consider only the longitudinal magnetic flux density which is better
constrained.

\subsection{Velocity and inclination discontinuities}

Gradients and/or discontinuities in velocity and magnetic field cause
asymmetries in Stokes $V$ profiles. The high-$\beta$ simulation (P06)
demonstrated how velocity gradients from shocking acoustic waves cause
a pattern of time-varying Stokes $V$ asymmetries in the Ca lines.
 
To study the effect of discontinuities we use the inversion code in
synthesis mode to produce a set of Stokes profiles from atmospheres
with velocity discontinuities of different sizes (1, 3 and 5 km/s) and
at different locations (between log$(\tau)=[-5.5, 0]$. We then compute Stokes $V$ asymmetries for the
profiles. The asymmetries are defined as in \cite{MartinezPillet+others1997}. The amplitude asymmetry of a Stokes V
profile is given by

\begin{eqnarray*}
\sigma_{\mathrm{a}}= \frac {a_b-a_r}{a_b+a_r},
\end{eqnarray*}
where $a_b$ and $a_r$ are the unsigned extrema of the blue and red
lobes of the Stokes $V$ profile.

The area asymmetry of a Stokes $V$ profile is given by:

\begin{eqnarray*} 
\label{eqn:area}
\sigma_{\mathrm{A}}=s\frac{\int_{\lambda_0}^{\lambda_1} V(\lambda)d\lambda}{\int_{\lambda_0}^{\lambda_1} |V(\lambda)|d\lambda},
\end{eqnarray*}
where $s$ is the sign of the blue lobe.

\begin{figure*}
\epsscale{.8}
\plotone{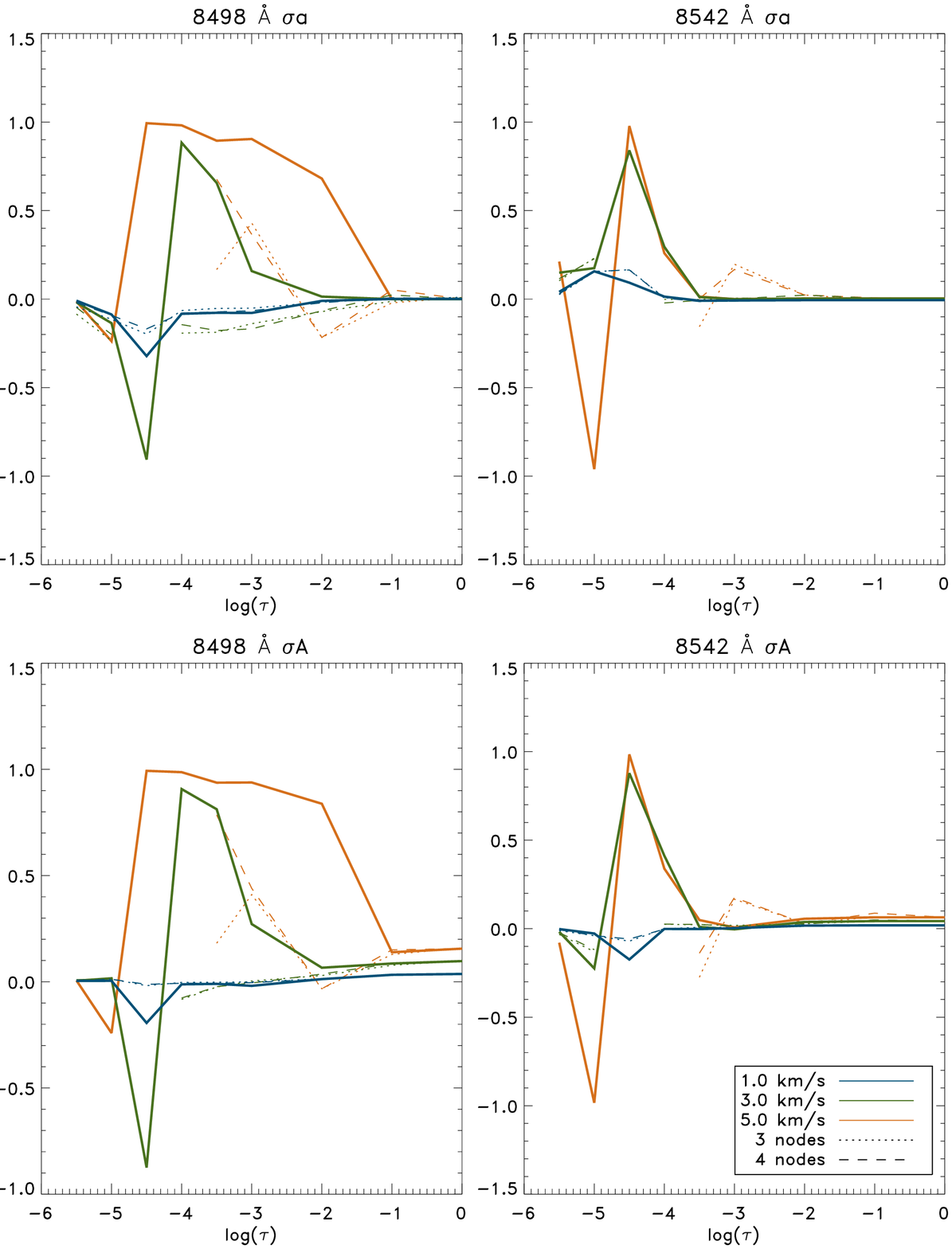}
\caption{
\label{fig:vel-discont1}
Asymmetries in Stokes $V$ profiles caused by discontinuities in velocity. Upper row: Stokes V amplitude asymmetry in 8498 \AA\ (left) and 8542 \AA\ (right) as a function of the location of the discontinuity. Green, blue and red solid lines are discontinuities of 1, 3 and 5 km/s, respectively. Dotted and dashed lines are asymmetries from inversion fits for 3 and 4 velocity nodes, respectively. Color coding is the same as for the solid lines. Where no inversion value is plotted, no satisfactory fit was found. Lower row: As upper row except for Stokes V area asymmetry. }
\end{figure*}

In figure ~\ref{fig:vel-discont1} the results are shown for the velocity discontinuities. For the sign of the velocity we use the standard
astrophysical notation where positive values indicate red-shifts. The sign and amplitude of the Stokes $V$ asymmetries depend both on
the location and size of the discontinuity. Note that both positive
and negative asymmetries are present even though all velocities are
positive and in all cases the
photosphere has a smaller velocity. The overall patterns in the 8498 \AA\ and 8542 \AA\ lines
are not identical, or even all that similar.

To see if the inversion code can handle velocity discontinuities we
invert the profiles (Stokes $I$ and $V$ of the two Ca lines and the
two photospheric Fe lines) using either 3 or 4 nodes in
velocity. Asymmetries computed from the inversion fits are shown in
dashed and dotted lines in figure \ref{fig:vel-discont1}. In most
cases the fits do not reproduce the asymmetries. However, the overall
qualitative shape of the line profiles is in most cases
satisfactory. Increasing the number of velocity nodes to 10 does not significantly improve the results.

Also discontinuities in inclination cause complex asymmetry patterns in the lines. The effects of a discontinuity on Stokes
$V$ asymmetries are not easily predictable since they depend strongly
on both the size and location of the discontinuity. Furthermore, the
discontinuities can cause the two Ca lines, with relatively close
formation heights and similar radiative transfer, to exhibit very
different asymmetries. Also the area and amplitude asymmetries can be
of opposite signs. A combination of discontinuities or strong
gradients in velocity and inclination will be even more
difficult to disentangle.


\section{Inversions of quiet Sun observations}
\label{sec:obs}
We use the non-LTE inversion code to invert quiet Sun observations of the
\ion{Ca}{2} 8498 \AA\ and 8542 \AA\ lines and \ion{Fe}{1} 8497 \AA\
and 8538 \AA\ lines taken with the the Spectro-Polarimeter for
INfrared and Optical Regions ({\em SPINOR},
\citealt{Socas-Navarro+others2006a}) at the Dunn Solar telescope,
Sacramento Peak Observatory. The raster scan was obtained in a quiet Sun
region near disk center at S17.3 W32.1 on May 19, 2005 at 14:14
UT. The upper part of the slit was positioned above internetwork and lower part above an enhanced network patch consisting of a
single polarity. For a detailed description of the
observations see P07.

The inversion is performed in 2 cycles. In the first cycle both Stokes
$I$ and $V$ are weighted equally and there are 4, 1 and 3 nodes in
temperature, magnetic field strength and velocity, respectively. In
the second cycle more weight (factor 500) is put on the Stokes $V$
profile and the number of nodes are increased to 7, 3 and 4
(temperature, magnetic field strength and velocity). A node for
macroturbulence is also added. The second cycle allows the code to fine-tune the solution
obtained in cycle 1. In both cycles the magnetic field inclination and
azimuth are forced to zero. 5 different initial models are used for
each inversion and the best fit is chosen based on the smallest
$\chi^2$ value.

A 2-component atmosphere is needed to fit most of the line profiles satisfactorily. The filling factor is a free
parameter and one of the atmospheric components is a non-magnetic external atmosphere which is the same for all pixels. The non-magnetic external atmosphere is obtained by inverting an average  non-magnetic
profile using the same number of nodes in velocity and temperature as
in the main inversions, but with no nodes in magnetic field and a zero
magnetic field in the initial models. The filling factor is forced to
be unity. The
non-magnetic profile is an average of intensity profiles in all pixels
where the 8498 \AA\ Stokes $V$ amplitude is below
$7\times10^{-3}$. Because of atomic polarization, Stokes $Q$ and $U$
are given no weight in the inversions. Consequently, we do not have
enough information to resolve both atmospheric components, hence the
non-magnetic atmosphere is set to be constant.

Since the telluric lines present are weak, they
cannot be used for a precise absolute wavelength calibration. Instead,
the velocities are calibrated so that the mean photospheric velocity
in ``non-magnetic'' regions (pixels where the photospheric magnetic field strength is
less than 300 G) is
set to zero.
 
\subsection{Goodness of fits}

\begin{figure*}
\epsscale{1}
\plotone{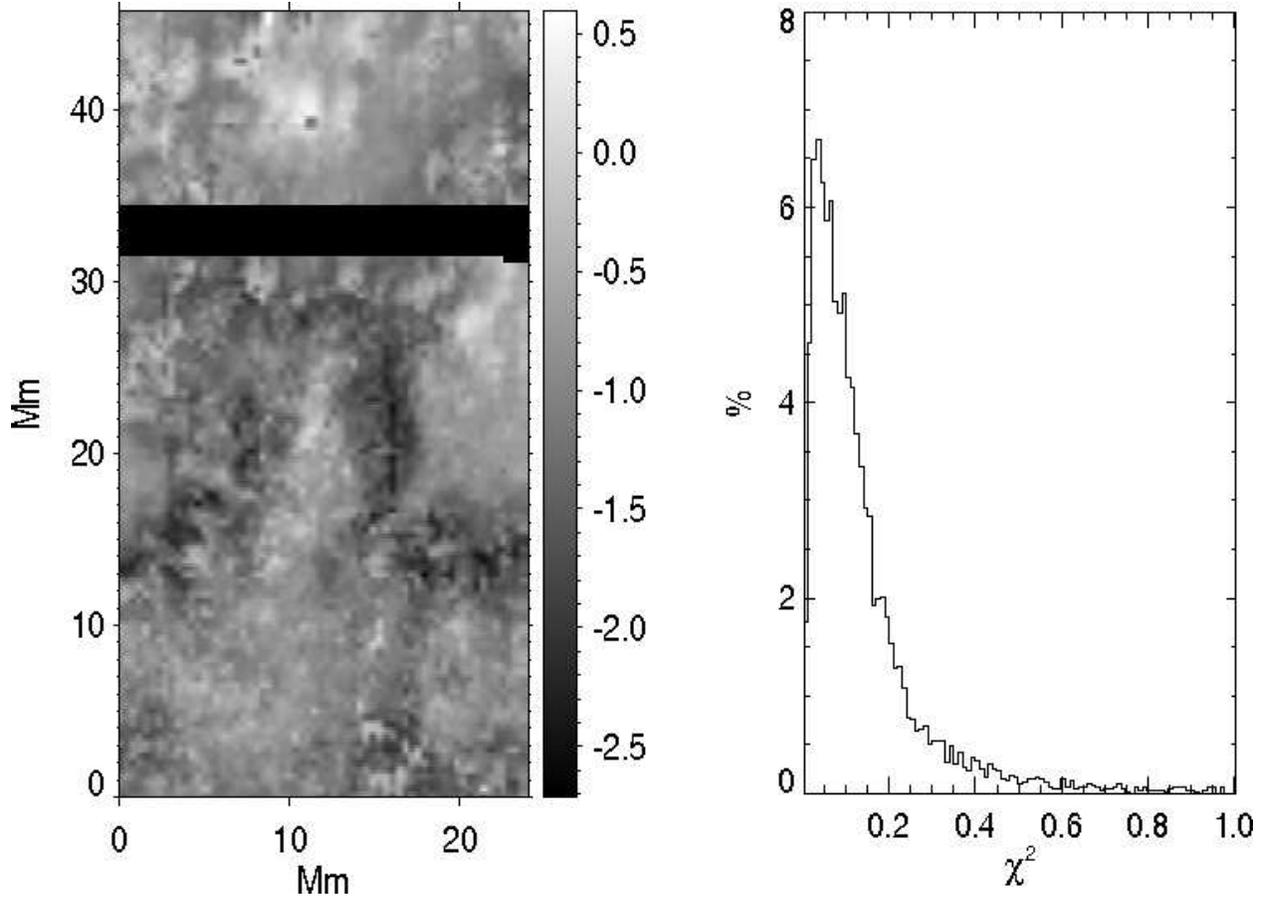}
\caption{
\label{fig:chi}
$\chi^2$ of the inversion fits.  right: image of the spatial distribution of $log_{10}(\chi^2)$. The black strip covers the hairline and the close-by pixels that are affected by it. Left: histogram of the $\chi^2$ values. }
\end{figure*}

\begin{figure*}
\epsscale{0.8}
\plotone{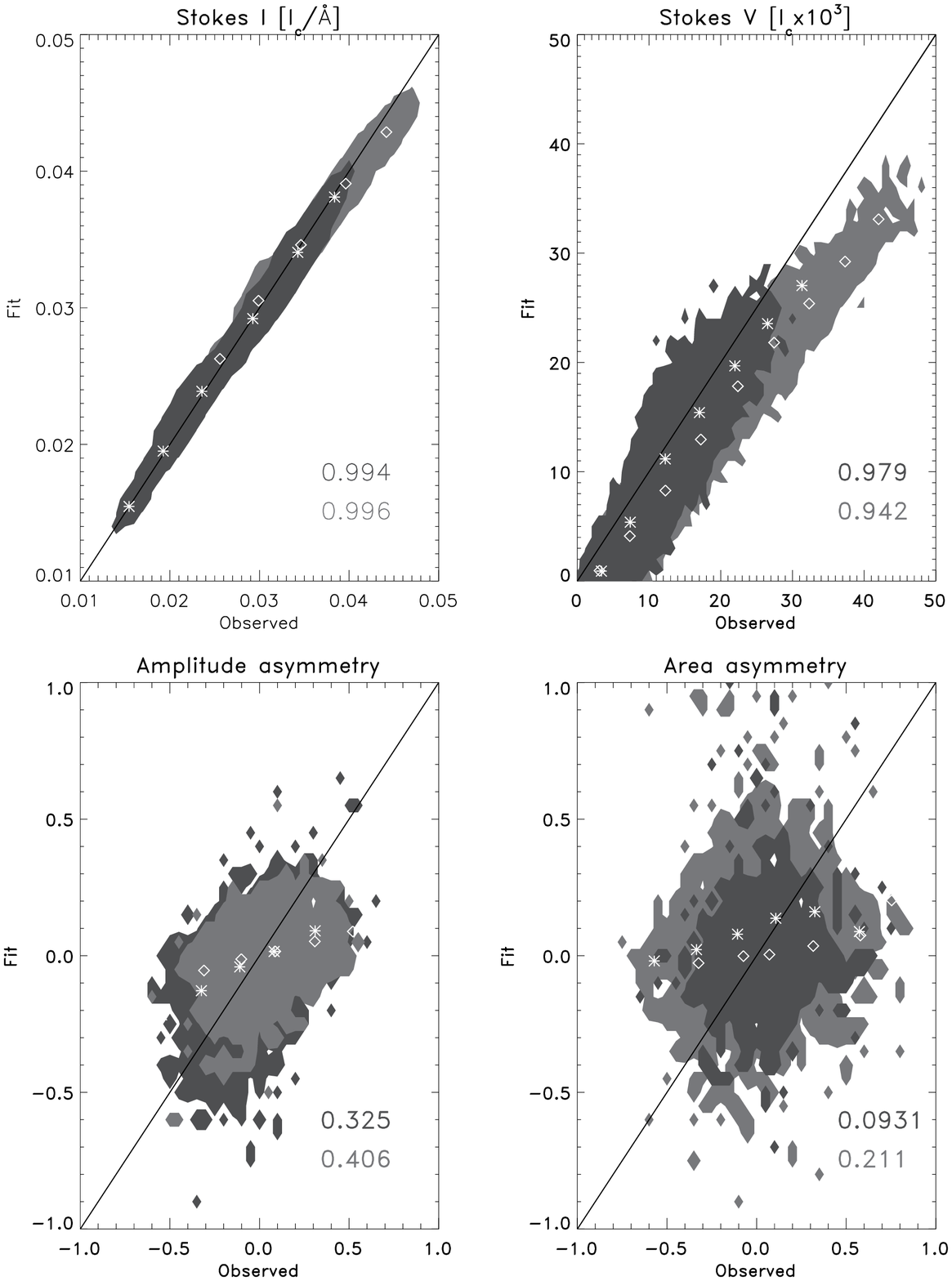}
\caption{
\label{fig:proxy}
Scatter plots of the observed and fitted 8498 \AA\ and 8542 \AA\ Stokes $I$ and $V$ amplitudes, and Stokes $V$ area and amplitude asymmetries. The over-plotted diamonds and asterisks show the mean for 8498 \AA\ and 8542 \AA\ lines, respectively. In the bottom right corners of the figures are cross correlation coefficients for the observed and fitted values. Darker gray is 8498 \AA\ and lighter gray 8542 \AA. }
\end{figure*}

$\chi^2$ was used as a proxy for the goodness of the fits produced by
the inversions (figure \ref{fig:chi}). In general, the $\chi^2$-values are below 0.2. Figure \ref{fig:proxy} illustrates how well the inversions retrieve
quantities used to describe the line profiles. The inversion code can
retrieve the amplitudes of both Stokes $I$ (integrated intensity in a
0.75 \AA\ wide wavelength band around the line core zero wavelength
normalized to the band width, $I_c/$\AA) and $V$ very well, but the
Stokes $V$ asymmetries are not reproduced. The 8498 \AA\
area asymmetry is especially problematic. Since the two Ca lines have very
different statistics in terms of the area asymmetries (P07) and, as shown in the previous section, discontinuities in
velocity and/or inclination cause complex asymmetry patterns, the
failure is not all that surprising. The only quantity that shows a
correlation with $\chi^2$ is Stokes $V$ amplitude. This may be
explained by it being easier to fit only the intensity profiles instead of fitting both the
Stokes $I$ and $V$ profiles. Also it is possible, if not likely, that
the magnetic elements are sub-resolution and the pixel consists of a
magnetic component and an external non-magnetic component. It is
uncertain how well the average external non-magnetic atmosphere describes the
non-magnetic component. No clear correlation is found between $\chi^2$
and Stokes $V$ asymmetries.

\subsection{Spatial patterns}

Since the inversions have the filling factor as a free parameter, the
results need to be weighted in respect to how much of the light from
the pixel originates from the magnetic component and how much from the
external non-magnetic component. Since this atmosphere is assumed the
same for all pixels and only the filling factor varies, for the
following discussion, only pixels
with a filling factor larger than 15 \% are considered without further
restrictions. For pixels with filling factors less than 15 \%, only
those with 8498 \AA\ Stokes $V$ amplitude above 0.02 $I_c$ are
included. The Stokes $V$ amplitude threshold was chosen to be the
amplitude above which the standard deviation of $\chi^2$ in models producing good
fits in a given pixel is the same for pixels with filling factors less
than 15\% and pixels with filling factors greater than 15 \%. For pixels with filling factors less than
15 \% and 8498 \AA\ Stokes $V$ amplitude below 0.02, the standard
deviation of the inverted atmospheres is significantly larger, and
they are omitted from the analysis.

\begin{figure*}
\epsscale{1}
\plotone{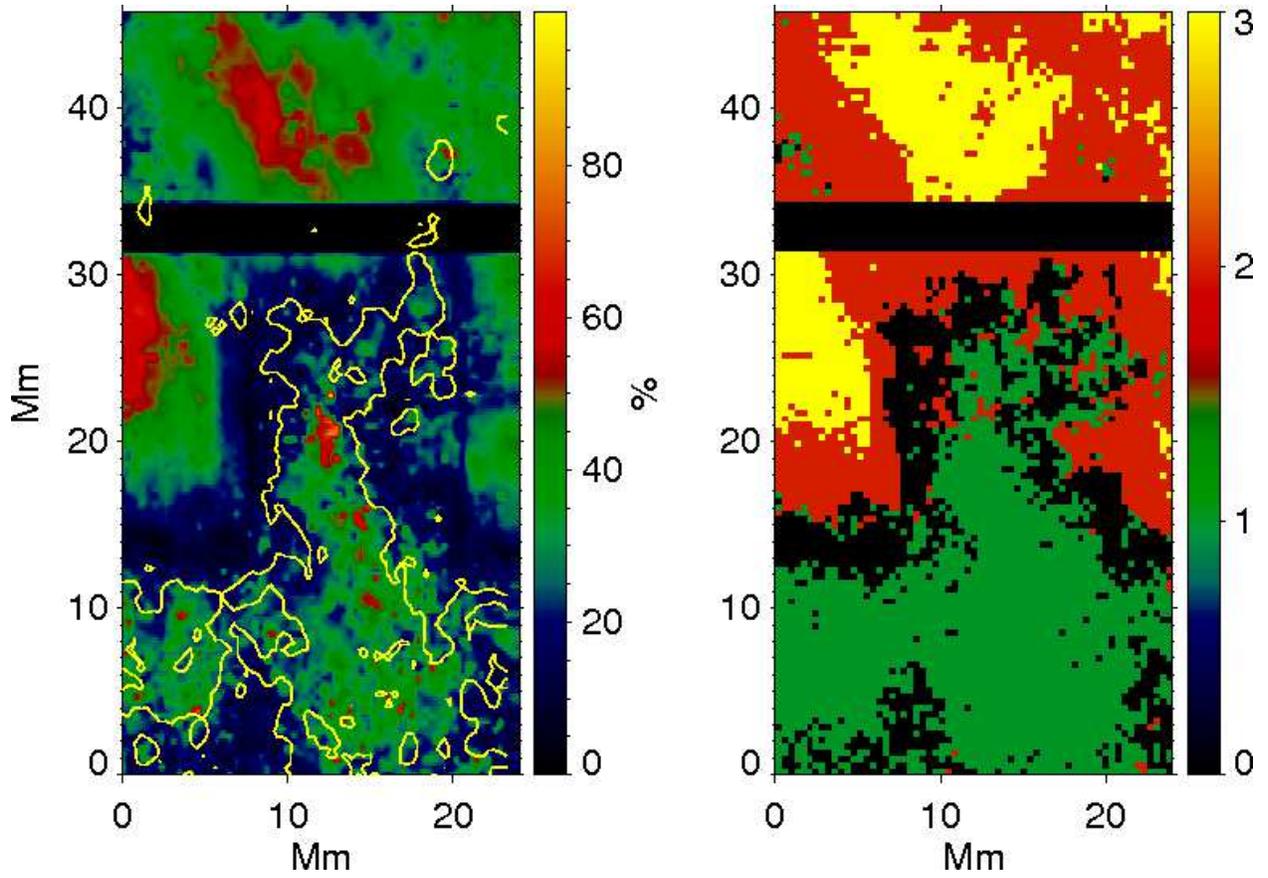}
\caption{
\label{fig:alfas}
Right: filling factors for the inversions. The contours are for 75 G photospheric longitudinal magnetic flux. Left: mask based on chromospheric temperature used to define the enhanced network, low-flux network and internetwork. }
\end{figure*}

Note that the filling factors are largest in the network and the very quiet internetwork whereas the
areas surrounding the network have very small filling factors (figure
\ref{fig:alfas}). This illustrates two points: the network is best
described with a 2-component scenario. Second, since the internetwork
has no signal in Stokes $V$ the code can combine the non-magnetic
profile with a significant contribution from an additional
(essentially also non-magnetic) atmospheric component resulting in a
2-component non-magnetic atmosphere. In the regions by the network,
the Stokes $V$ amplitude is in general small, but not negligible. Since the Stokes $V$ signal is weighted more by the inversion
code in
such cases the non-magnetic component is described solely by the
external non-magnetic atmosphere even though it is the dominant
component.  Since there is little signal in Stokes $V$ in these pixels, consequently also the
filling factors will be small.

\begin{figure*}
\epsscale{1}
\plotone{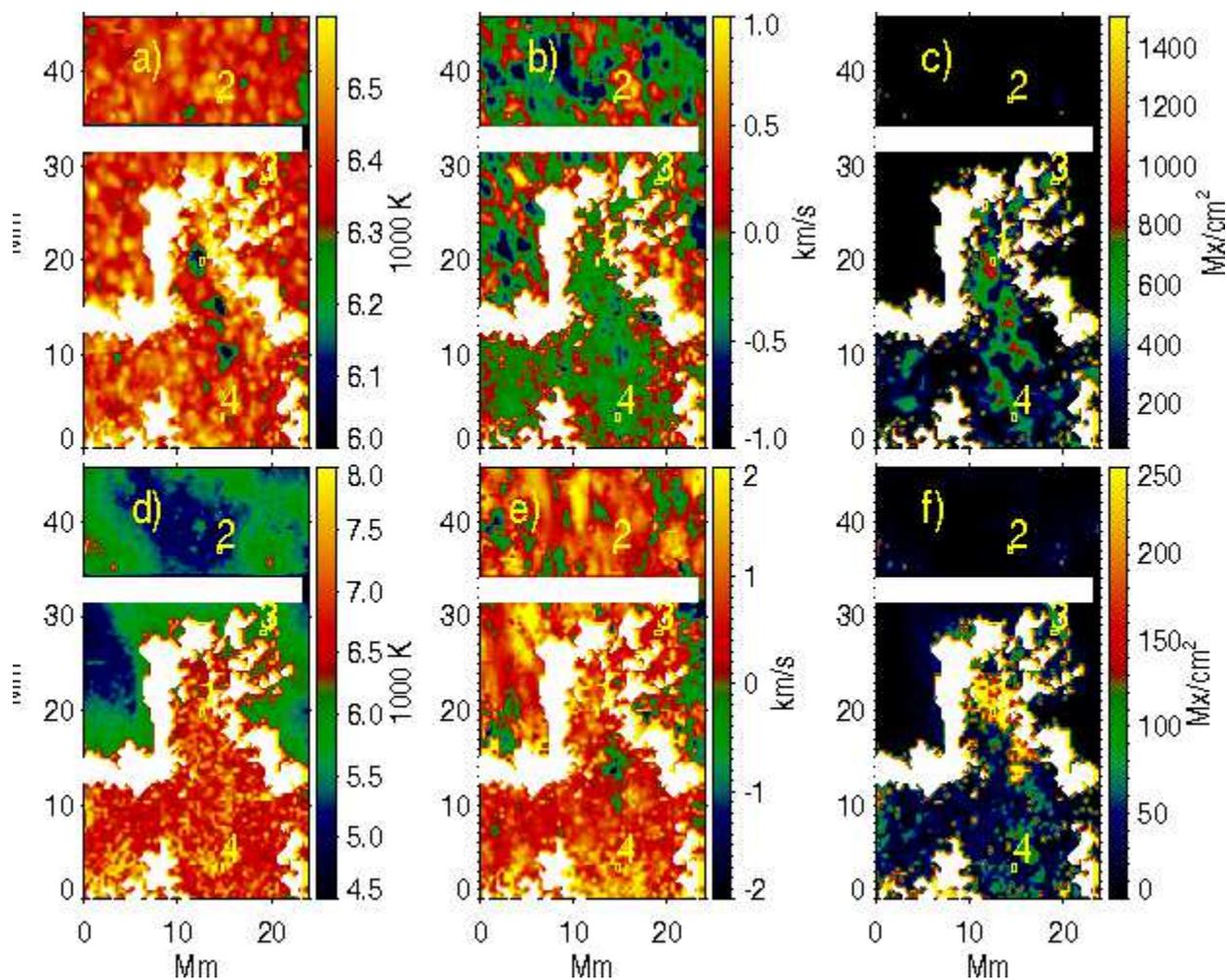}
\caption{
\label{fig:atmos-image}
{\it a)} Photospheric temperature (at log($(\tau)$)=0).  {\it b)} Photospheric velocity (at log($(\tau)$)=-1). {\it c)} Photospheric longitudinal magnetic flux. (at log($(\tau)$)=-1) {\it d)-f)} as {\it a)-c)} but for the chromosphere (at log($(\tau)$)=-5, -5.5 and -5). For explanation of choice of pixels, see text. Numbers indicate locations of pixels discussed in more detail in section \ref{sec:ind}.}
\end{figure*}

The photospheric longitudinal magnetic flux density (at log$(\tau)$=-1, hereafter referred to as the magnetic flux, panel {\it c)} in figure \ref{fig:atmos-image}) has a
fragmented pattern with distinct edges. The strongest flux, about 1200
Mx/cm$^2$, is in the centers of flux ``bundles''. There are no
isolated points of strong flux. This is due to both the photospheric
lines being weak and thus not very sensitive to magnetic fields, and
to the spatial resolution of the observations. Since individual flux
tube size is below the spatial resolution, only regions with a concentration of several flux
tubes appear strong. The strongest flux in the chromosphere
(at log$(\tau)$=-5, panel {\it f)} in figure \ref{fig:atmos-image}) does not lie directly above the photospheric
flux concentrations, and not all of the strong flux seen in the
photosphere reaches the chromosphere. In fact, the strongest flux
concentrations are significantly weaker higher up, and the total area
covered by strong flux is smaller. The magnetic network is more
confined in the photosphere. In the chromosphere the field
has expanded to fill more space causing the more diffuse appearance of the network.

The photospheric temperature (at log$(\tau)$=0, panel {\it a)} in figure \ref{fig:atmos-image}) is significantly
cooler in the strong flux concentrations. To maintain a pressure
balance between the interior and exterior of the flux tube, the gas
pressure inside the tube must be smaller than outside leading to a reduced density inside the flux tube. Also the convective energy transport in the tube is suppressed by the magnetic field. If the diameter of the flux tube (or concentration of tubes) is large enough the lateral radiative flux from the hotter surroundings does not reach the tube center, and it remains cooler (and appears as a dark feature in intensity images). Hints of a granulation pattern are visible in
photospheric temperature: the cell centers are
hotter than the edges. The network is well visible in the
chromospheric temperature (at log$(\tau)$=-5, panel {\it d)} in figure \ref{fig:atmos-image}), and is clearly hotter than the
internetwork. The locations of the hottest pixels are not correlated
with the strongest flux. Instead the hot pixels form a fragmented
pattern that covers most of the region where flux is seen. The network
boundaries are visible as a region of intermediate temperature,
clearly distinct from both the network and internetwork. The
intermediate temperature region extends to form a network like pattern
in the mostly flux-free region (hence hereafter this region
will be referred to as low-flux network). There is no clear indication
of the low-flux network in photospheric temperature or
magnetic flux.

Since no absolute wavelength calibration was done, defining the zero velocity is not exact and velocities close to zero
should not be taken as definite measures. In the photospheric
internetwork patches of negative and positive velocities are
present (at log$(\tau)$=-1, panel {\it b)} in figure \ref{fig:atmos-image}). The network is dominated
by negative velocities. Some positive velocities are also seen, though
they are located closer to the edges and are not as large in amplitude
as in the internetwork. There are large scale
patterns of positive velocities, interlaced with significantly smaller
patches of negative velocities, in the chromospheric internetwork (at log$(\tau)$=-5.5, panel {\it e)} in figure \ref{fig:atmos-image}). The chromospheric network has
mostly positive velocities, though above one of the strong flux
concentrations a patch with a large negative velocity is seen. The
overall velocity pattern in the chromosphere is more fragmented in the network, and also the velocity amplitudes vary less than in the internetwork. This may be due to the underlying magnetic field
``compartmentalizing'' the plasma. There is little correlation between photospheric and chromospheric velocities.

\subsection{Histograms of atmospheres}

\begin{figure*}
\epsscale{1}
\plotone{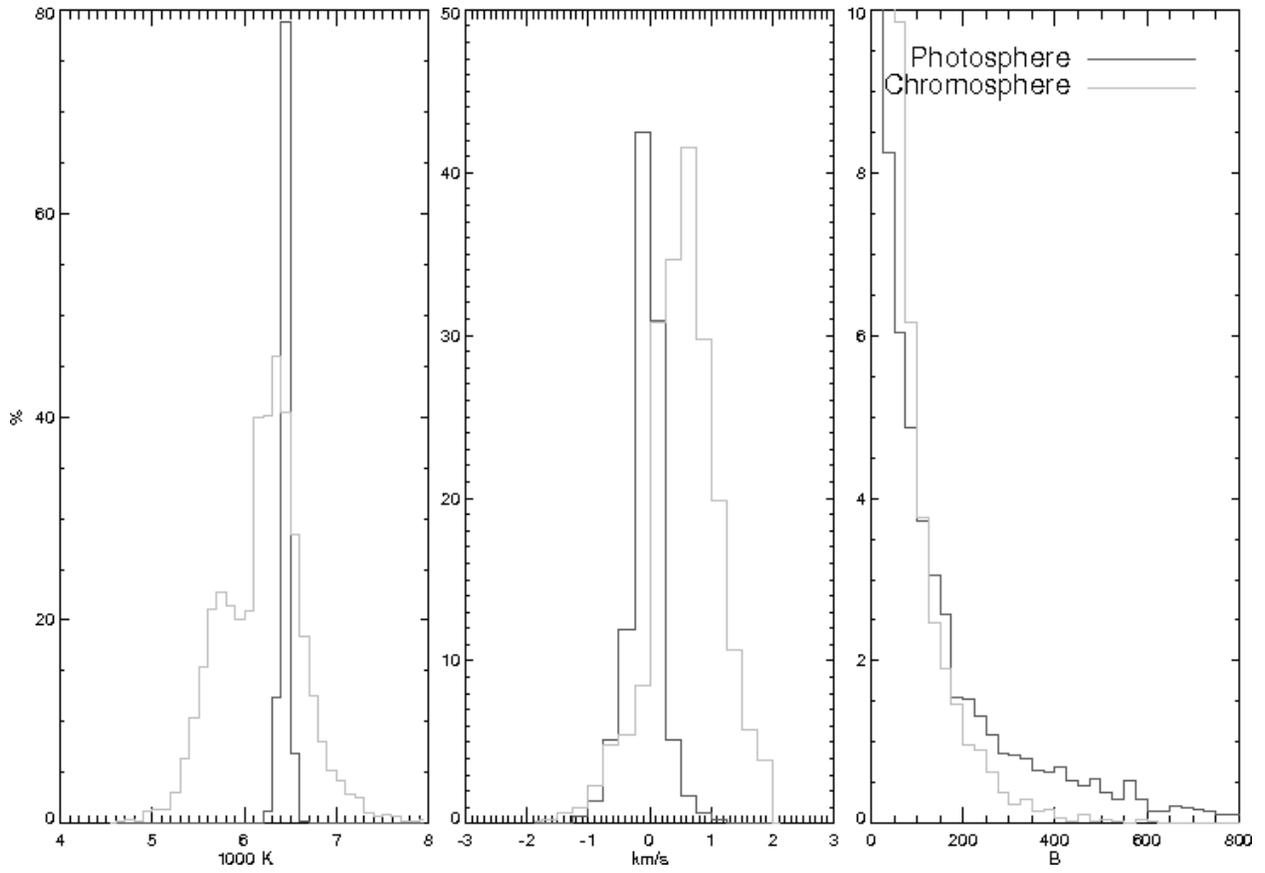}
\caption{
\label{fig:hists}
Left: Histograms of photospheric (dark gray line) and chromospheric (light gray line) temperature (left), velocity (middle) and longitudinal magnetic flux (right).}
\end{figure*}

In figure \ref{fig:hists} are shown histograms of the
photospheric and chromospheric temperature, magnetic flux and
velocity. The photospheric temperature histogram is
narrower than the chromospheric one. It has a single peak at 6400
K, which is also the temperature in the photosphere of the external
non-magnetic atmosphere. The chromospheric temperature histogram has
2 peaks: the first one is at 5800 K, and corresponds to the low-flux
network. The coolest pixels, below 5400 K, are in the
internetwork. The second peak, corresponding to the hot network
pixels, at 6400 K, is narrower but has an extended tail to high
temperatures. This is higher than in the external non-magnetic
atmosphere (6260 K).

The photospheric velocity histogram is centered at 0 km/s,
which is also by definition the photospheric velocity in the external
non-magnetic atmosphere and in pixels with a photospheric magnetic flux less than 300 G.  The chromospheric velocity histogram peaks
at a positive values: 0.5 km/s. It is significantly wider than the photospheric histogram.

The longitudinal magnetic flux in both the photosphere and chromosphere has
exponentially decaying histograms. The flux in the photosphere decays
slower than in the chromosphere. Neither histogram has peaks
corresponding to the network flux. 

\begin{figure*}
\epsscale{0.9}
\plotone{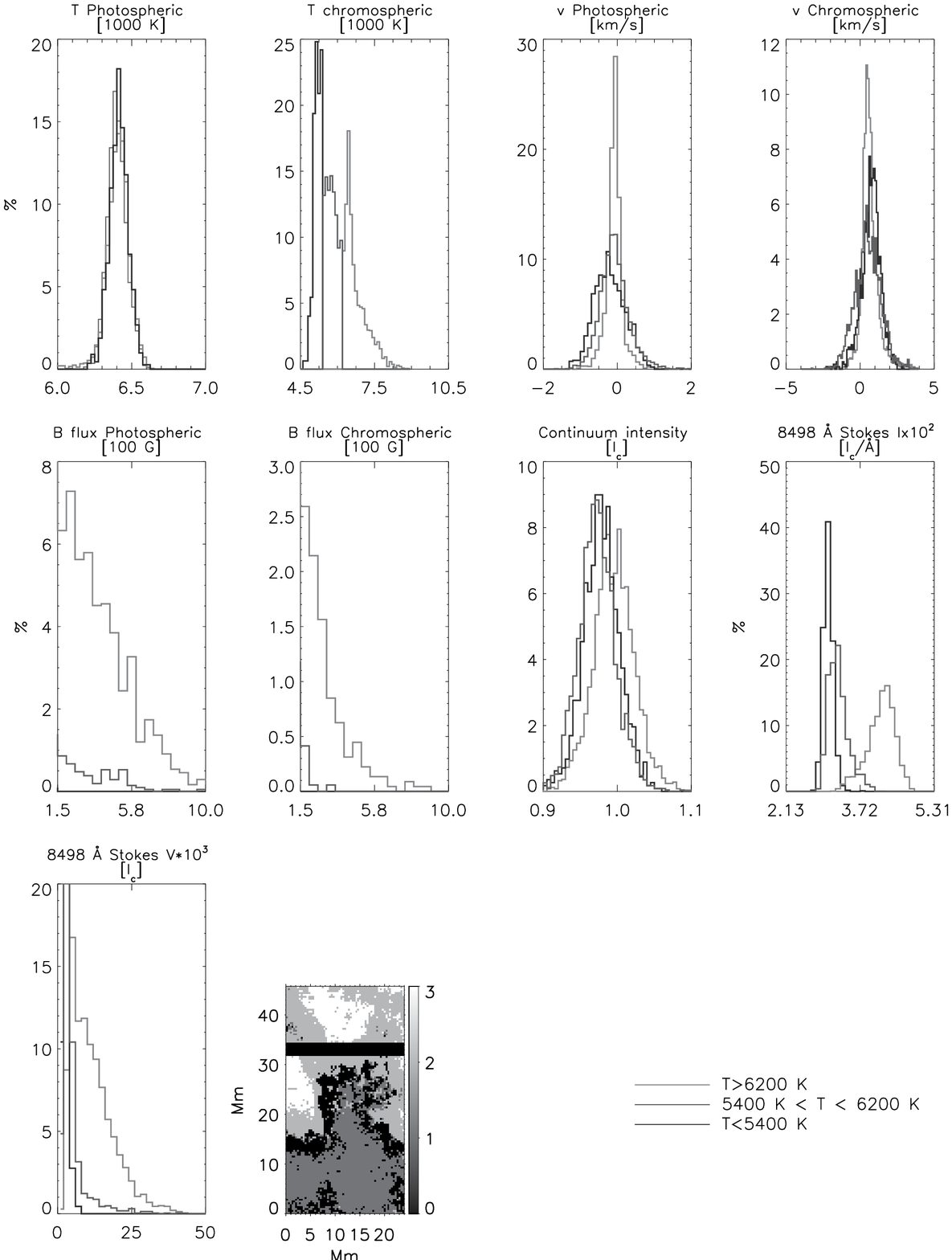}
\caption{
\label{fig:nw-inw}
Histograms of photospheric and chromospheric temperatures, velocities and magnetic flux, continuum intensities and Ca line amplitudes in the internetwork, low-flux network and network. Image shows the mask used to identify the pixels: dark gray is enhanced network, light gray low-flux network and white internetwork. The black regions are pixels not included in the analysis.}
\end{figure*}

There are three distinct regions in the chromospheric temperature: the enhanced
network, low-flux network and internetwork. We define a
mask based on the temperature: the
internetwork is where the chromospheric temperature is below 5400 K, low-flux network where the temperature
is between 5400 K and 6200 K and network where the temperature is
above 6200 K. The mask clearly separates the map into three distinct
regions (shown in figures \ref{fig:alfas} and \ref{fig:nw-inw} where the three colors mark the three regions). Histograms
illustrate similarities and differences between the three regions and the line
profiles in them. The photospheric
temperature distributions are nearly identical in all three
regions: the network, except for the cool strong flux concentrations,
is not visible in photospheric temperature.

The photospheric velocity histograms are centered around zero or close to
zero. The chromospheric histograms peak at positive values with the internetwork having the largest red-shift: it peaks at 1 km/s. The enhanced network regions have the narrowest velocity histograms both in the photosphere and in the chromosphere. This is consistent with the magnetic field suppressing convection. However, the bottoms of the histograms in the enhanced network are roughly as wide as in the internetwork histograms. There is no significant difference between the internetwork and low-flux network histograms in the photosphere. In the chromosphere, however, the low-flux network histogram is significantly wider. 

As expected, the magnetic flux is by far largest in the network. No
flux is seen in the internetwork. (Since the photospheric lines are
weak and the seeing conditions during the observing sequence were not
optimal, no internetwork fields are seen.) The low-flux network
has some, though very little when
compared to the network, magnetic flux.

The network is clearly brightest
of the three regions when viewed in continuum intensity. No difference is seen between the internetwork and low-flux
network. In the Ca line intensities the low-flux network is somewhat brighter than the
internetwork, but still clearly darker than the network. The Stokes
$V$ amplitudes of the Ca lines further illustrate that the
low-flux network is magnetic: the amplitudes are larger in
the low-flux region than in the internetwork. In general, the differences
between the three regions, especially between the internetwork and
low-flux network, become more pronounced with height.

\begin{figure*}
\epsscale{1}
\plotone{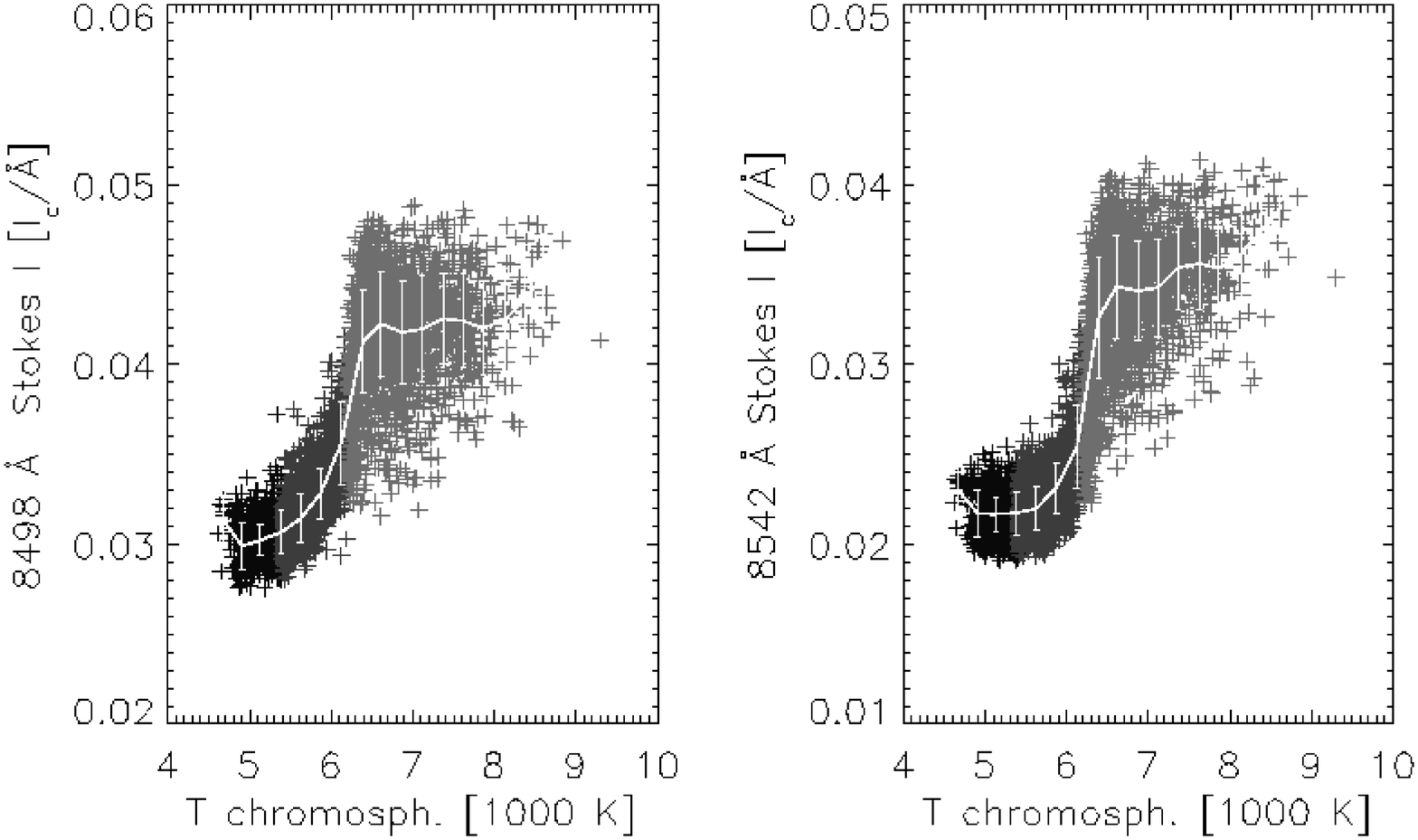}
\caption{
\label{fig:i-vs-flux}
8498 \AA\ and 8542 \AA\ Stokes $I$ amplitudes versus chromospheric temperature. The network, low-flux network and internetwork are shown in different shades of gray: the dark is the internetwork, medium dark the low-flux network and light the enhanced network. }
\end{figure*}

The dependence of the Ca lines' Stokes $I$ amplitudes on the
chromospheric temperature is shown in figure \ref{fig:i-vs-flux}. Both
lines show a saturation in intensity at high temperatures, and a
minimum level intensity at low temperatures. The intensities show a
strong dependence on the temperature only in the intermediate temperature
range. Implications on the use of proxies based on spectral lines formed in non-LTE are discussed in section
\ref{sec:disc}. Scatter plots of chromospheric temperature and
longitudinal magnetic flux, both photospheric and chromospheric,
show some signs of saturation at high fluxes, but the spread is large
and the number of data points with large magnetic flux is not high
enough for the saturation to be determined as statistically significant.

\subsection{Individual atmospheres}
\label{sec:ind}

In the following, individual atmospheres representative of different
regions seen in the map are discussed. Locations of the regions are marked in figure \ref{fig:atmos-image}.

\begin{figure*}
\epsscale{0.8}
\plotone{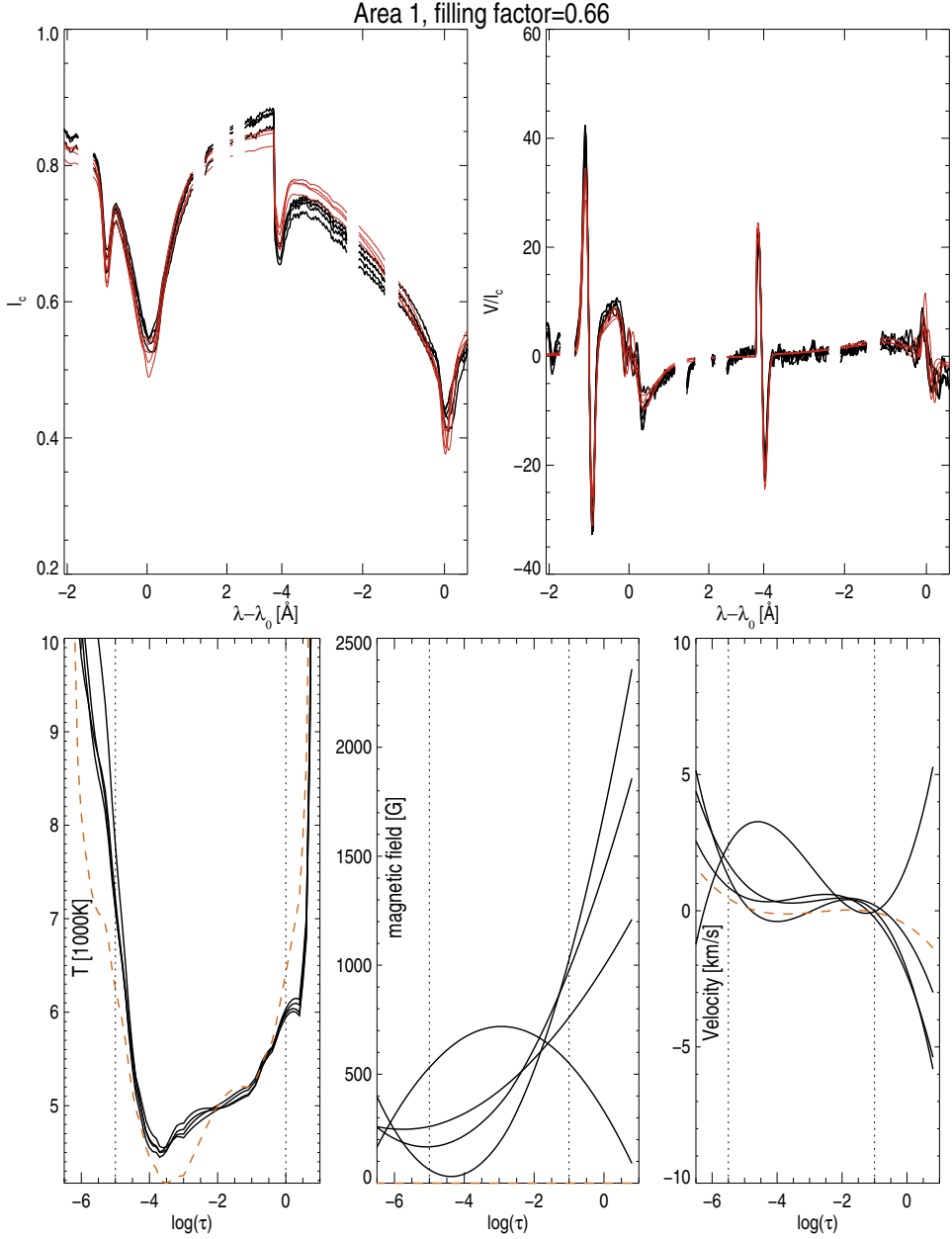}
\caption{Upper row Stokes $I$ and $V$ profiles for area 1 (for location see figure \ref{fig:atmos-image}). Observed profiles and inversion fits are shown in black and red lines, respectively. Lower row: Temperature (left), longitudinal magnetic flux (middle) and velocity (right) as a function of optical depth. The dashed red lines in the temperature and velocity panels are for the external non-magnetic atmosphere. The vertical dotted lines show where the inversion code results are most reliable.
\label{fig:a1}
}
\end{figure*}

Line profiles and atmospheres for pixels belonging to area 1 are shown in
figure \ref{fig:a1}. Area 1 is located in a strong photospheric flux-bundle. It is also clearly visible in photospheric
temperature. The line profiles are all fairly symmetric and similar to
one another (except for one asymmetric 8498
\AA\ Stokes $V$ profile. It corresponds to the atmosphere that deviates most from the mean
in both the magnetic field and velocity). The average filling factor of the pixels
is 0.66. The Stokes $V$ profiles of
the photospheric iron lines are identical to one another indicating that the photospheric magnetic field is
fairly homogeneous in area 1. Photospheric temperature in the atmosphere is ~430 K cooler than in the external
non-magnetic photosphere. The chromosphere is significantly
hotter, by 1000 K. The
longitudinal magnetic flux
is of the order of 1 kG in the photosphere and decays by a factor of 5
before reaching the chromosphere. This is already seen in the images of magnetic flux where the strong
photospheric flux concentrations are not visible in the
chromosphere. The velocity is zero in the photosphere and positive in the chromosphere.

\begin{figure*}
\epsscale{0.8}
\plotone{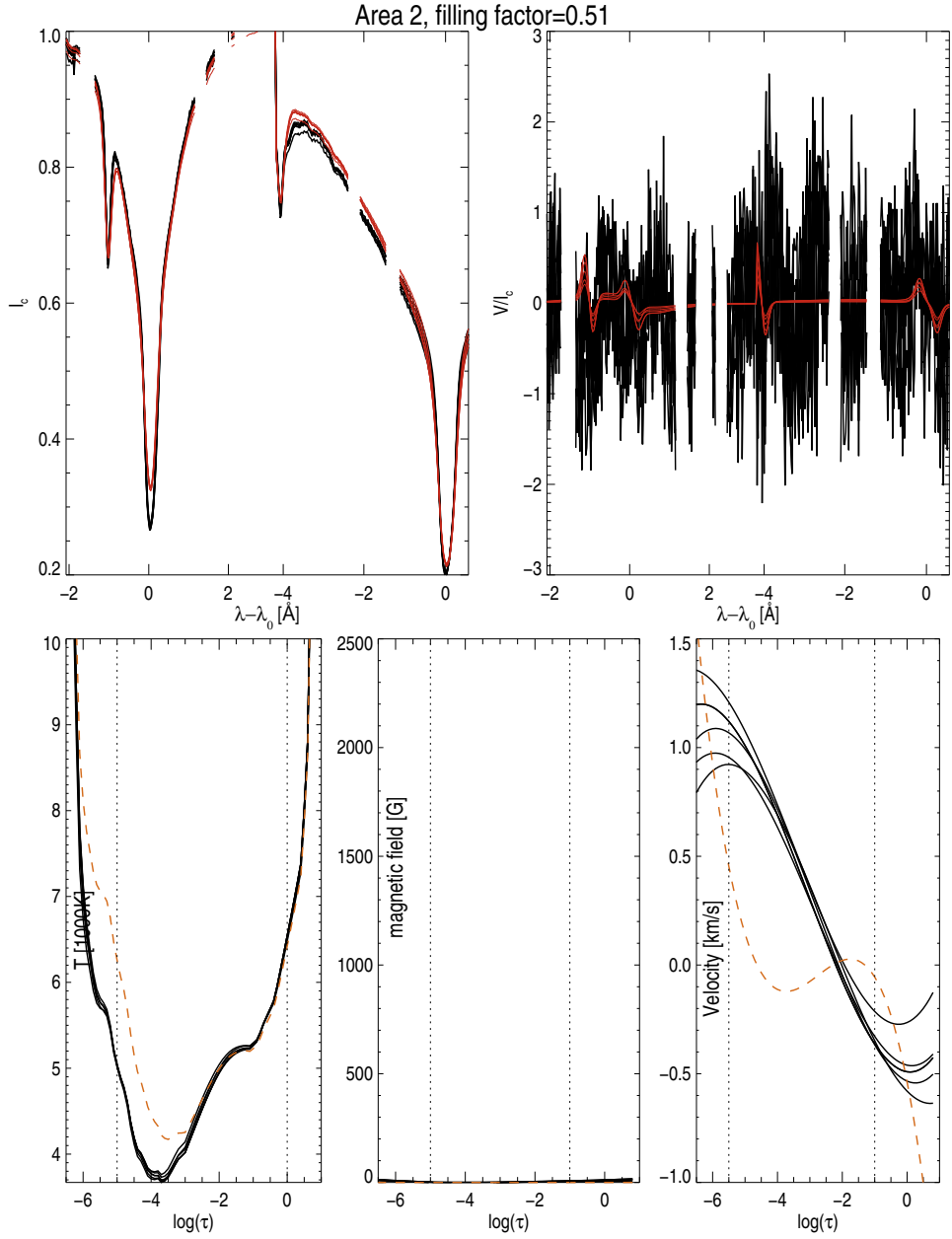}
\caption{As figure \ref{fig:a1} but for area 2.
\label{fig:a2}
}
\end{figure*}

Area 2 (figure \ref{fig:a2}) is located deep in the internetwork. The
Stokes $I$ profiles are all nearly identical: deep, narrow and very
symmetric. There is no signal above noise in Stokes $V$, and
consequently, the filling factor (0.51) only describes how much of the
average non-magnetic profile is present in the line profiles. The temperature
in the photosphere is nearly identical to the external non-magnetic
atmosphere, but the chromosphere is significantly cooler. Because the pixels for the non-magnetic profile were chosen based on
magnetic signal and not on intensity, pixels in the low-flux
network (which are hotter than the internetwork) were included. The inversion code returns a flux that is less
than 20 Mx/cm$^2$, and the fitted Stokes $V$ profiles are well below
noise level. The velocity profile differs from the non-magnetic both
in the photosphere and chromosphere. This is expected since the non-magnetic profile is an average of many pixels with
different phases, i.e., all but net shifts should be eliminated from
it. In general, there is very little scatter between the model
atmospheres indicating that the region is fairly homogeneous, both in
terms of thermal and dynamic properties.

\begin{figure*}
\epsscale{0.8}
\plotone{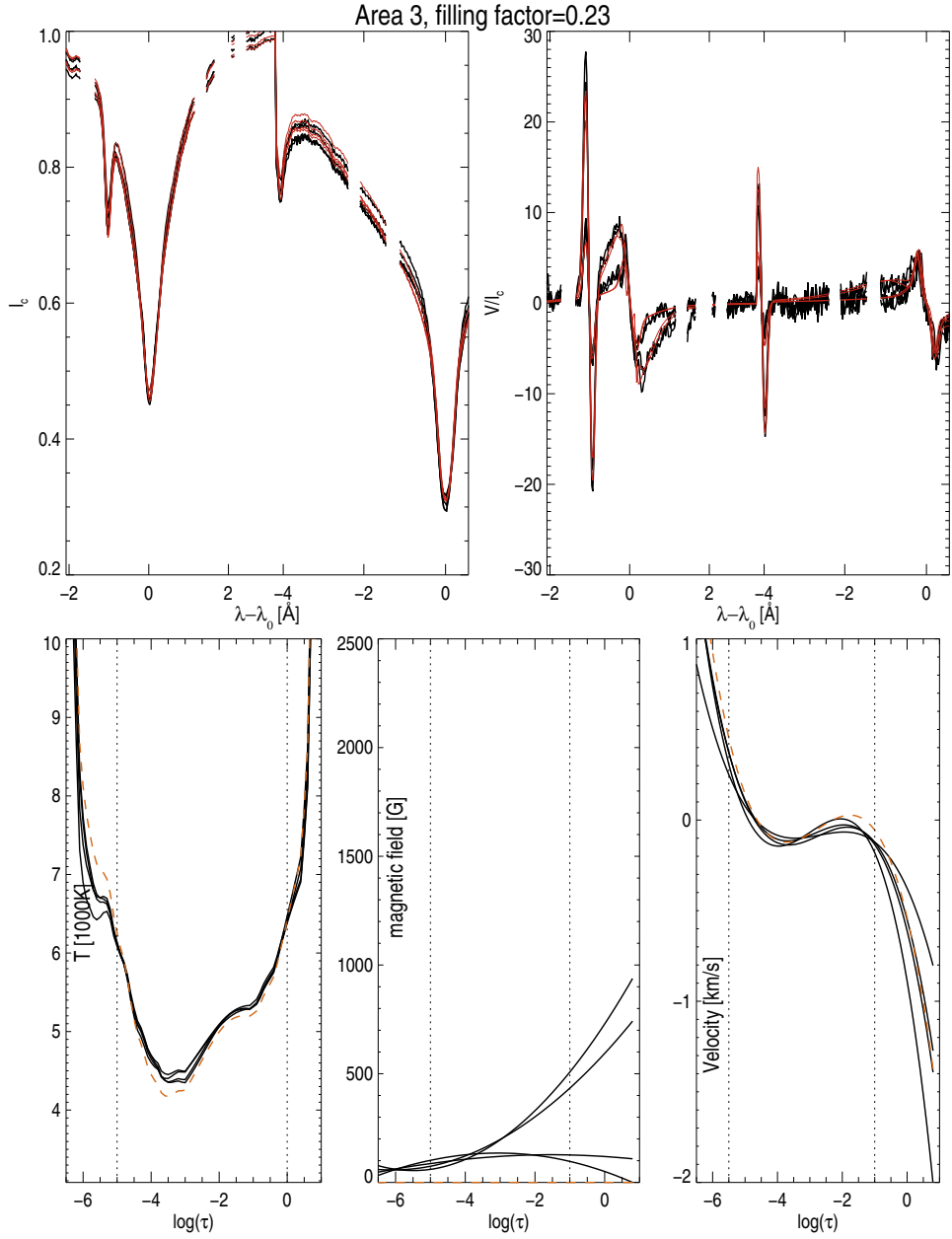}
\caption{As figure \ref{fig:a1} but for area 3.
\label{fig:a3}
}
\end{figure*}

Pixels in area 3 (figure \ref{fig:a3}) are located in the network, but outside
of the strong flux concentrations. The Stokes $I$ and $V$ profiles are
fairly symmetric, but only the 8542 \AA\ line has nearly identical Stokes
$V$ profiles in all pixels. The other 3 lines show two sets of Stokes
$V$ amplitudes: one set significantly larger than the other. This is
especially clear in the photospheric iron lines. The 8498 \AA\ line
has a set of Stokes V profiles with strong signal in the line
wings and a set with little signal in the wings and also a
slightly smaller amplitude. The temperature in the photosphere is roughly the same as in the external non-magnetic
atmosphere: the coolest photosphere in area 3 is only 50 K cooler. This is a very small difference when
compared to area 1 where the Stokes $V$ profiles had similar shapes
(albeit somewhat larger amplitudes) as here. In contrast to area 1 where the
chromosphere is significantly hotter than the external non-magnetic
atmosphere, in area 3 the chromosphere is
cooler, by roughly 150 K. Above log$(\tau)=-6$ the
temperature decreases even further. This is
likely a consequence of spline interpolation when the code is adjusting the
temperature in the lower parts of the atmosphere. Since the
lines are not sensitive to conditions above ~$log(\tau)=-6$ this
peculiar temperature profile has no effect on the fits. As expected based on the two sets of Stokes $V$
amplitudes, there are two sets of magnetic fluxes in the
photosphere. The spread of fluxes in the chromosphere is significantly smaller. Both
the photospheric and chromospheric velocities are lower than in area
1. The velocities
are slightly negative in the photosphere and positive in the chromosphere.The models are in fairly
good agreement with one another. Since the code cannot retrieve sizes of velocity gradients, this
does not rule out that the differences in the 8498 \AA\ and 8542 \AA\
profile shapes are not caused by large gradients or discontinuities in
velocity (or inclination).

\begin{figure*}
\epsscale{0.8}
\plotone{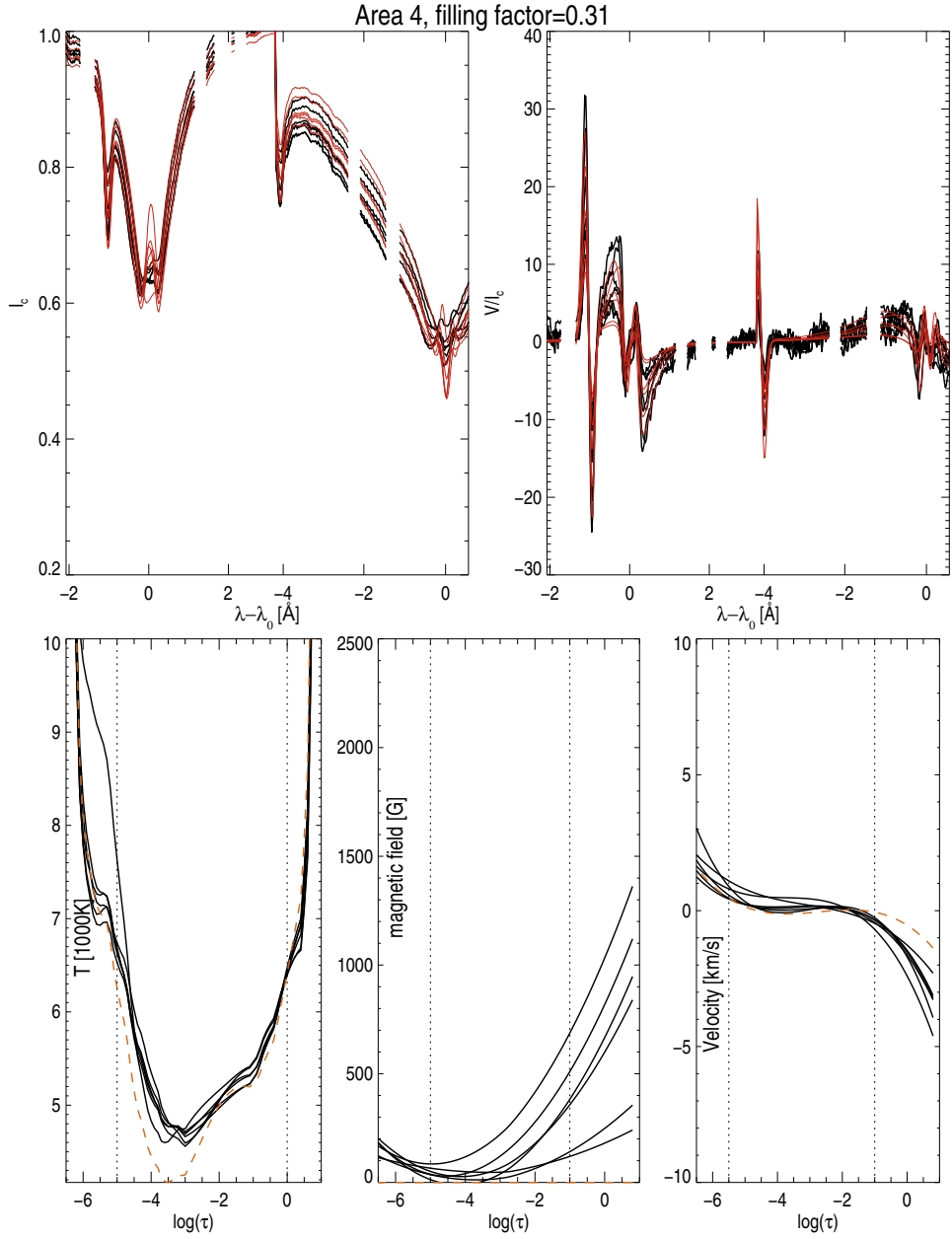}
\caption{As figure \ref{fig:a1} but for area 4.
\label{fig:a4}
}
\end{figure*}

Area 4 (figure \ref{fig:a4}) is also located in the network, but even farther from the strong photospheric magnetic flux concentrations than area 3. The chromospheric Stokes $I$ profiles show
self-reversals which are not seen in area 3. The self-reversals are stronger on the red side in the 8542 \AA\ line and in
most cases also in the 8498 \AA\ line. The chromospheric Stokes $V$ profiles all
show varying degrees of asymmetry where the red sides tend to have
multiple components. The inversions reproduce the general shapes of
the profiles, but not the amplitudes of the self-reversals and the
emission features in Stokes $V$. Since
the external non-magnetic atmosphere is based on an average profile of
both the map and time series, anything leading to asymmetries in Stokes $I$
has to be incorporated into the magnetic component. The photospheric
temperature is not very different from the external non-magnetic
atmosphere but the chromosphere is clearly hotter.

\section{Discussion}
\label{sec:disc}

Two points need to be kept in mind when interpreting the inversion results. Firstly, the sensitivity of the inversions is limited to specific
heights. Since the Fe lines and Ca line wings are formed in the photosphere, and only the Ca line cores are chromospheric, the sensitivity range is larger in the
photosphere than in the chromosphere. Because the
inversion code does not include treatment of atomic polarization (and
consequently Stokes $Q$ and $U$ cannot be used in the weak field
regime), it is not possible to resolve the ambiguity between the
magnetic field strength and inclination, instead only the longitudinal
magnetic flux density can be retrieved.

Secondly, discontinuities in velocity and magnetic field inclination
result in complicated Stokes $V$ asymmetry patterns. The responses to
discontinuities in the two Ca lines are not always similar. Also the area and amplitude asymmetries
can respond in different ways. The
inversion code cannot reproduce the discontinuities. Such discontinuities may explain the
observed Stokes $V$ area and amplitude asymmetry histograms (P07): the 8542 \AA\ area asymmetry histogram
peaks at a positive value whereas all other histograms (8542 \AA\ amplitude asymmetry, 8498 \AA\ area and amplitude asymmetry) peak at a
negative value. Based on the high-$\beta$ simulations it seems
unlikely that the difference is caused by shocking acoustic waves alone. 
The current inversions are not a good diagnostic for discontinuities
or strong localized gradients, especially since both velocity and inclination discontinuities can be
present. Since we do not include linear polarization in the inversions, and since the asymmetries are not reproduced by the inversions, there is information in the spectra which is not used by the current inversion code. A more suitable inversion code, especifically designed for the inhomogeneities of the chromosphere, is needed to solve the problem.

Despite the limitations, these non-LTE inversions remain a useful tool: observed Stokes
$I$ and $V$ amplitudes are reproduced successfully
even if asymmetries in Stokes $V$ are not.

\subsection{Magnetic field}

The magnetic contrast of the network and surroundings is significantly reduced
in the chromosphere, and there are clear indications of the network expanding with height. These indications include the spatial distribution of Stokes $V$ amplitudes in the Ca and Fe lines, the appearance of the magnetograms (figure 2.\ in P07) and spatial distributions of temperature and magnetic flux in the photosphere and chromosphere. The diffusive appearance and spreading of the
network in the chromosphere was already noted by
\cite{Harvey+Hall1971} who attributed it to the magnetic field
becoming more horizontal with height. This is in agreement with our
data, and especially with the clearly more diffuse and larger appearance of the network
in the 8542 \AA\ magnetogram when compared to the 8498 \AA\ magnetogram. However, since we do not have
information of the field inclination, we cannot conclude this for
sure.

Not all of the strong magnetic flux concentrations in the photosphere
penetrate into the chromosphere. One possible cause is the expansion of
the field with height. This would reduce the longitudinal flux and increase
the amount of horizontal field on the expense of vertical
field. Another explanation is that some of the flux returns back to
the photosphere. \cite{Schrivjer+Title2003},
\cite{Jendersie+Peter2006} and \cite{Aiouaz+Rast2006} have explored
the connection of the network magnetic field to the corona, and shown
that the expansion of the network is affected by magnetic fields of
opposite polarity in the network and internetwork. Unfortunately, our
observations of the photospheric magnetic field are greatly limited by
the weakness of the iron lines. There is no indication of opposite
polarity magnetic field in our data, but as pointed out by the
above-mentioned authors, {\it sub-resolution} magnetic fields in the
internetwork play a crucial role in determining what fraction of the
network flux survives to the corona and what fraction returns to the
photosphere forming loops at various heights. The low-flux network
is a clear indication of there being a substantial amount of magnetic
flux not captured by the photospheric iron lines.

Based on the inspection of the individual atmospheres the photosphere appears more homogeneous, especially in terms of
magnetic flux. There is more small-scale spatial variation in the
chromosphere. The photosphere is unlikely to have large gradients or
discontinuities, and the photospheric radiation is in
LTE. Consequently, it is easier to find a model to fit the
observations in the photosphere and the spread of atmospheric models
will be smaller there. However, since the small-scale chromospheric
spatial patterns are coherent over more than just one pixel, and because
we see spatial variations on small scales in the line profiles as
well, it is likely that the inhomogeneous appearance of the chromosphere is real.

\subsection{Dynamics}
 
Velocities in the chromosphere, especially in most of the network,
are predominantly red-shifted in relation to the photosphere. The
chromospheric network velocity patterns are fragmented and large
velocities are more often seen outside the network. The latter is also
true in the photosphere where the difference is even more
pronounced. A similar finding is presented in \cite{Rezaei+others2007} where the fraction of
large velocities (determined from photospheric Stokes $V$ profiles) is
larger in the internetwork than in the network. In the current data, there is no difference between the
low-flux network and internetwork in terms of photospheric
velocities, but in the chromosphere the distribution of velocities is
slightly wider in the low-flux regions. The strongest magnetic fields suppress oscillations, although
large asymmetries  are also seen in network line profiles. In general, the largest asymmetries are located at network boundaries. Also increasing Stokes $V$ asymmetries are correlated with decreasing magnetic flux. The asymmetries are in most cases at least partly due to
emission features in the line profiles, i.e. they are also likely to be related
to the temperature structure, not just velocity and magnetic field gradients.  The difference between
the velocity histograms in the low-flux network and internetwork
indicates that magnetic fields that do not play a visibly important
role in the photosphere, become important higher up.

Recent work by e.g., \cite{Jefferies+others2006} and
\cite{Vecchio+others2007}, has tied together the
magnetic topology and the observed oscillatory properties in the chromosphere. The network edges being more dynamic is
consistent with the picture presented by \cite{Jefferies+others2006}. In it inclined magnetic fields channel low-frequency waves into the
chromosphere. \cite{Jefferies+others2006} also point out that the low-frequency waves are
intermittent owing to the continual changes caused by the motion of
the convective cells and by the ``magnetic carpet'' (i.e., small
magnetic dipoles created in the internetwork and transported to the network where they interact with existing flux). Reconnection events would also probably lead to asymmetries in line profiles. In our data the
network clearly expands with height, and the difference between the
8498 \AA\ and 8542 \AA\ magnetograms is striking, indicating that
conditions described above (i.e., inclined fields and more dynamic network boundaries) are likely to exist. Furthermore, since the
photospheric driver has most of its power at periods of 5 min, the
driver of chromospheric oscillations in regions with inclined fields
is stronger than in the internetwork where the 5 minute period
waves are filtered out. This was pointed out by \cite{DePontieu+others2007}
in connection to the dynamic behavior of fibrils. All the above is consistent
with the network, and especially the boundaries, being more dynamic
and exhibiting more self-reversals and profile asymmetries than the
internetwork.

\subsection{Temperature proxies and heating of the magnetic chromosphere}

\label{sec:heating}
Figure \ref{fig:i-vs-flux} shows that the intensity of the Ca II IR triplet lines
saturates at high chromospheric temperatures. A similar saturation
effect is seen if the magnetic flux is plotted instead of the
temperature. In \cite{Rezaei+others2007} a similar plot, but for the
Ca H intensity and photospheric flux, is interpreted as a
saturation limit for the magnetic heating (i.e., the heating as a
function of field strength or filling factor levels off at high
values). The inversions show that the leveling off in the Ca II IR triplet lines is, at least partly,
a radiative transfer effect, and does not necessarily imply that the
magnetic heating has an upper limit that depends on the field
strength or filling factor. The data do not conclusively show this not
to be the case either. (In fact, there are some signs of a saturation effect
in scatter plots of magnetic flux and chromospheric temperature.)
Also a lower limit on chromospheric emission, similar to that seen in
\cite{Rezaei+others2007}, is seen in our data, but no lower limit on
the chromospheric temperature is found. The \ion{Ca}{2} IR triplet line intensities are
linearly dependent on temperature only in the range
between roughly 5800 and 6500 K. Interpreting proxies based on lines
formed outside of LTE is not straight-forward, and the effects of
radiative transfer need to be considered in detail when inferring physical properties, such as velocity and temperature, from proxies.

The network patch appears on large scales fairly
uniform in temperature, and the hottest pixels do not
always coincide with strongest photospheric or chromospheric flux. Nor do
they coincide with the largest filling factors. Based on this the
relationship between heating and magnetic flux does not appear to be
linear. However, since the low-flux network is cooler, the amount of flux
is clearly a factor in the heating. We see signs of a saturation effect in
the heating (flux as a function of chromospheric temperature) as
proposed by \cite{Rezaei+others2007} and references therein. The
number of data points is not, however, large enough for making definite
conclusions. Further observations and inversions, preferably
simultaneously with strong photospheric lines, are needed.

\section{Final remarks}
\label{sec:conc}
Because of non-LTE it is important to take into account radiative
transfer effects when interpreting proxies based on line
profiles. This is why non-LTE inversions are an invaluable tool: with them it is possible to infer macroscopic physical quantities without resorting to proxies. The inversions presented in this work
further highlight the physical differences between the dynamics and structure of the chromospheric
network and internetwork. The chromosphere is divided into three regions
characterized best by chromospheric temperature, but also visible in other chromospheric parameters. The fundamental differences
between the three regions are most likely related to magnetic fields.

There are limitations to what can be done with current
non-LTE inversion codes, and forward modeling remains a crucial
part in understanding the dynamics of the magnetic chromosphere. We
plan to make simultaneous observations of the Ca II IR triplet lines and
the \ion{He}{1} 10830 \AA\ lines to further validate the inversions,
to obtain a more 3-dimensional view of the magnetic chromosphere, and
to study the connection between thermal properties of the chromosphere (magnetic heating) and magnetic field topology.


\begin{thebibliography}{30}
\expandafter\ifx\csname natexlab\endcsname\relax\def\natexlab#1{#1}\fi

\bibitem[{{Aiouaz} \& {Rast}(2006)}]{Aiouaz+Rast2006}
{Aiouaz}, T. \& {Rast}, M.~P. 2006, \apjl, 647, L183

\bibitem[{{Centeno} {et~al.}(2005){Centeno}, {Socas-Navarro}, {Collados}, \&
  {Trujillo Bueno}}]{Centeno+others2005}
{Centeno}, R., {Socas-Navarro}, H., {Collados}, M., \& {Trujillo Bueno}, J.
  2005, \apj, 635, 670

\bibitem[{{De Pontieu} {et~al.}(2004){De Pontieu}, {Erd{\'e}lyi}, \&
  {James}}]{DePontieu+others2004}
{De Pontieu}, B., {Erd{\'e}lyi}, R., \& {James}, S.~P. 2004, \nat, 430, 536

\bibitem[{{De Pontieu} {et~al.}(2007){De Pontieu}, {Hansteen}, {Rouppe van der
  Voort}, {van Noort}, \& {Carlsson}}]{DePontieu+others2007}
{De Pontieu}, B., {Hansteen}, V.~H., {Rouppe van der Voort}, L., {van Noort},
  M., \& {Carlsson}, M. 2007, \apj, 655, 624

\bibitem[{Gabriel(1976)}]{Gabriel1976}
Gabriel, A. 1976, Phil Trans. Royal Soc. Lond., 281, 339

\bibitem[{{Hansteen} {et~al.}(2006){Hansteen}, {De Pontieu}, {Rouppe van der
  Voort}, {van Noort}, \& {Carlsson}}]{Hansteen+others2006}
{Hansteen}, V.~H., {De Pontieu}, B., {Rouppe van der Voort}, L., {van Noort},
  M., \& {Carlsson}, M. 2006, \apjl, 647, L73

\bibitem[{{Harvey} \& {Hall}(1971)}]{Harvey+Hall1971}
{Harvey}, J. \& {Hall}, D. 1971, in IAU Symp. 43: Solar Magnetic Fields, ed.
  R.~{Howard}, 279--+

\bibitem[{{Jefferies} {et~al.}(2006){Jefferies}, {McIntosh}, {Armstrong},
  {Bogdan}, {Cacciani}, \& {Fleck}}]{Jefferies+others2006}
{Jefferies}, S.~M., {McIntosh}, S.~W., {Armstrong}, J.~D., {Bogdan}, T.~J.,
  {Cacciani}, A., \& {Fleck}, B. 2006, in ESA SP-624: Proceedings of SOHO
  18/GONG 2006/HELAS I, Beyond the spherical Sun

\bibitem[{{Jendersie} \& {Peter}(2006)}]{Jendersie+Peter2006}
{Jendersie}, S. \& {Peter}, H. 2006, \aap, 460, 901

\bibitem[{Judge {et~al.}(2001)Judge, Casini, Tomczyk, Edwards, \&
  Francis}]{Judge+others2001}
Judge, P., Casini, R., Tomczyk, S., Edwards, D.~P., \& Francis, E. 2001,
  Coronal Magnetogmetry: a feasibility study, Tech. Rep. NCAR/TN-446-STR,
  National Center for Atmospheric Research

\bibitem[{{Lagg} {et~al.}(2004){Lagg}, {Woch}, {Krupp}, \&
  {Solanki}}]{Lagg+others2004}
{Lagg}, A., {Woch}, J., {Krupp}, N., \& {Solanki}, S.~K. 2004, \aap, 414, 1109

\bibitem[{{Mart{\'{\i}}nez Pillet}(1997)}]{MartinezPillet+others1997}
{Mart{\'{\i}}nez Pillet}, V. et~al. 1997, \apj, 474, 810

\bibitem[{{McIntosh} {et~al.}(2003){McIntosh}, {Fleck}, \&
  {Judge}}]{McIntosh+Fleck+Judge2003}
{McIntosh}, S.~W., {Fleck}, B., \& {Judge}, P.~G. 2003, \aap, 405, 769

\bibitem[{{McIntosh} \& {Judge}(2001)}]{McIntosh+Judge2001}
{McIntosh}, S.~W. \& {Judge}, P.~G. 2001, Astrophys.\ J., 561, 420

\bibitem[{{Merenda} {et~al.}(2006){Merenda}, {Trujillo Bueno}, {Landi
  Degl'Innocenti}, \& {Collados}}]{Merenda+others2006}
{Merenda}, L., {Trujillo Bueno}, J., {Landi Degl'Innocenti}, E., \& {Collados},
  M. 2006, \apj, 642, 554

\bibitem[{{Pietarila} {et~al.}(2007){Pietarila}, {Socas-Navarro}, \&
  {Bogdan}}]{Pietarila+others2007a}
{Pietarila}, A., {Socas-Navarro}, H., \& {Bogdan}. 2007, in press

\bibitem[{{Pietarila} {et~al.}(2006){Pietarila}, {Socas-Navarro}, {Bogdan},
  {Carlsson}, \& {Stein}}]{Pietarila+others2006}
{Pietarila}, A., {Socas-Navarro}, H., {Bogdan}, T., {Carlsson}, M., \& {Stein},
  R.~F. 2006, \apj, 640, 1142

\bibitem[{{Rezaei} {et~al.}(2007){Rezaei} et~al.}]{Rezaei+others2007}
{Rezaei}, R. et~al. 2007, ArXiv Astrophysics e-prints

\bibitem[{{Rybicki} \& {Hummer}(1991)}]{Rybicki+Hummer1991}
{Rybicki}, G.~B. \& {Hummer}, D.~G. 1991, \aap, 245, 171

\bibitem[{{Sasso} \& {Solanki}(2006)}]{Sasso+Lagg+Solanki2006}
{Sasso}, C., L.~A. \& {Solanki}, S. 2006, A\&A, 456, 367

\bibitem[{{Schrijver} \& {Title}(2003)}]{Schrivjer+Title2003}
{Schrijver}, C.~J. \& {Title}, A.~M. 2003, \apjl, 597, L165

\bibitem[{{Socas-Navarro}(2005)}]{Socas-Navarro2005a}
{Socas-Navarro}, H. 2005, \apjl, 631, L167

\bibitem[{{Socas-Navarro} {et~al.}(1998){Socas-Navarro}, {Ruiz Cobo}, \&
  {Trujillo Bueno}}]{Socas-Navarro+RuizCobo+TrujilloBueno1998}
{Socas-Navarro}, H., {Ruiz Cobo}, B., \& {Trujillo Bueno}, J. 1998, \apj, 507,
  470

\bibitem[{{Socas-Navarro} \& {Trujillo
  Bueno}(1997)}]{Socas-Navarro+TrujilloBueno1997}
{Socas-Navarro}, H. \& {Trujillo Bueno}, J. 1997, \apj, 490, 383

\bibitem[{{Socas-Navarro} {et~al.}(2000{\natexlab{a}}){Socas-Navarro},
  {Trujillo Bueno}, \& {Ruiz Cobo}}]{Socas-Navarro+others2000b}
{Socas-Navarro}, H., {Trujillo Bueno}, J., \& {Ruiz Cobo}, B.
  2000{\natexlab{a}}, Science, 288, 1398

\bibitem[{{Socas-Navarro} {et~al.}(2000{\natexlab{b}}){Socas-Navarro},
  {Trujillo Bueno}, \& {Ruiz Cobo}}]{Socas-Navarro+TrujilloBueno+RuizCobo2000}
---. 2000{\natexlab{b}}, \apj, 530, 977

\bibitem[{{Socas-Navarro} {et~al.}(2006)}]{Socas-Navarro+others2006a}
{Socas-Navarro}, H. {et~al.} 2006, Solar Physics, 235, 55

\bibitem[{{Socas-Navarro} {et~al.}(2006)}]{Socas-Navarro+others2006b}
{Socas-Navarro}, H. et~al. 2006,
  Solar Physics, 235, 75

\bibitem[{{Trujillo Bueno} {et~al.}(2005){Trujillo Bueno}, {Merenda},
  {Centeno}, {Collados}, \& {Landi Degl'Innocenti}}]{TrujilloBueno+others2005}
{Trujillo Bueno}, J., {Merenda}, L., {Centeno}, R., {Collados}, M., \& {Landi
  Degl'Innocenti}, E. 2005, \apjl, 619, L191

\bibitem[{{Vecchio} {et~al.}(2007){Vecchio}, {Cauzzi}, {Reardon}, {Janssen}, \&
  {Rimmele}}]{Vecchio+others2007}
{Vecchio}, A., {Cauzzi}, G., {Reardon}, K.~P., {Janssen}, K., \& {Rimmele}, T.
  2007, \aap, 461, L1

\end{thebibliography}
\end{document}